\documentclass[12pt,preprint]{aastex}

\def\hi{\ifmmode {\rm H}\,{\sc i}~ \else H\,{\sc i}~\fi}
\def\kms{\rm\,km\,s^{-1}}

\def\chandra {{\it Chandra}}
\def\fuse {{\it FUSE}}
\def\cvi {\ion{C}{6}}
\def\nvii {\ion{N}{7}}
\def\neix {\ion{Ne}{9}}
\def\ovi {\ion{O}{6}}
\def\oviB {\ion{O}{6}$_{\rm B}$}
\def\oviN {\ion{O}{6}$_{\rm N}$}
\def\ovii {\ion{O}{7}}
\def\oviii {\ion{O}{8}}
\def\ovihv {\ion{O}{6}$_{\rm HV}$}
\def\novi {N_{\rm OVI}}
\def\novii {N_{\rm OVII}}
\def\noviii {N_{\rm OVIII}}

\def\kms    {~km~s$^{-1}$}

\slugcomment{Accepted to ApJ}

\shorttitle{Local X-ray Absorption Toward Mkn 279}
\shortauthors{Williams et al.}

\begin{document}


\title{\boldmath{\it Chandra} Detection of Local Warm--Hot Gas 
     Toward Markarian 279}


\author{Rik J. Williams\altaffilmark{1}, 
        Smita Mathur\altaffilmark{1},
        Fabrizio Nicastro\altaffilmark{2,3,4}}
\altaffiltext{1}{Department of Astronomy, The Ohio State University, 
                 140 West 18th Avenue, Columbus OH 43210, USA}
\altaffiltext{2}{Harvard--Smithsonian Center for Astrophysics, Cambridge, MA,
                 USA}
\altaffiltext{3}{Instituto de Astronom\'ia Universidad Aut\'onomica de
 M\'exico, Apartado Postal 70-264, Ciudad Universitaria, M\'exico, D.F.,
 CP 04510, M\'exico}
\altaffiltext{4}{Osservatorio Astronomico di Roma, Istituto Nazionale
di AstroFisica}
\email{williams, smita@astronomy.ohio-state.edu; nicastro@head.cfa.harvard.edu}



\begin{abstract}
We report the \chandra\ detection of \ovii\ K$\alpha$ absorption at $z=0$ in the
direction of the $z=0.03$ Seyfert 1 galaxy Mkn 279.  The high velocity cloud
Complex C lies along this line of sight, with \hi\ 21-cm emission and \ovi\
1032\AA\ absorption both observed at velocities of $\sim -150$\kms\ relative
to the local standard of rest.  We present an
improved method for placing limits on the Doppler parameter and column
density of a medium when only one unresolved line can be measured; this
method is applied to the \ovii\ absorption seen here, indicating that the
\ovii\ Doppler parameter is inconsistent with that of
any low--velocity (Galactic thick disk) or high--velocity \ovi\ (\ovihv) 
component.  
Direct association of the \ovii\ with the \ovihv\ is further ruled out by the
high temperatures required to produce the observed \ovihv/\ovii\ ratio
and the significant velocity difference between the \ovii\ and
\ovihv\ lines.  If the \ovii\ absorption is associated with a very broad,
undetected \ovi\ component, then the absorption must be broadened by primarily
nonthermal processes.  The large
velocity dispersion and possible slight redshift of the \ovii\ absorption
(as well as limits on the absorber's temperature and density) may be 
indicative of a local intergalactic medium origin, though absorption from
a hot, low--density Galactic corona cannot be ruled out.

\end{abstract}


\keywords{ intergalactic medium -- X-rays: galaxies: clusters -- cosmology:
 observations}


\section{Introduction}

The advent of high--resolution X-ray spectroscopy with 
the \chandra\ X-ray Observatory and XMM--Newton, as well as far--ultraviolet 
spectroscopy with the Far Ultraviolet Spectroscopic Explorer (FUSE) and 
the Space Telescope Imaging Spectrograph on HST, has shed a great deal of 
light on the 
local warm--hot intergalactic medium (WHIM).  Hydrodynamic simulations predict
that this tenuous web of $\sim 10^5-10^7$\,K gas should contain about half
of the baryons in the nearby universe \citep{cen99,dave01}, 
appearing as a forest of highly ionized metal absorption lines in 
high signal--to--noise X-ray spectra of background sources \citep{hellsten98,perna98}. 
Indeed, recent \chandra\ grating observations of this ``X-ray forest'' 
toward bright blazars have confirmed these predictions 
\citep[][Nicastro et al., in preparation]{nicastro05a,nicastro05b}.

In addition to the intervening absorption systems, similar metal absorption
lines (primarily \ovii) are observed at redshifts consistent with zero toward 
several background quasars, such as Mrk~421 \citep[][hereafter W05]{williams05},
PKS~2155-304 \citep[][Williams et al., in preparation]{nicastro02,fang02},
and 3C 273 \citep{fang03}; 
larger archival analyses have also been performed for various instruments
\citep[e.g.][]{mckernan04,fang06}.  This nearby 
absorption presents a  unique puzzle: since
no morphological or distance information is known about the X-ray absorption
(other than limited spatial data along the very few quasar pencil beams 
where it has been
detected), it is still unknown whether this warm--hot gas originates
within the Galactic halo or rather is part of the Local Group WHIM, or
a combination of the two.
While some \ovii\ absorption has been detected within 50\,kpc
of the Galaxy \citep{wang05}, this is unlikely to be uniformly distributed.
Additionally, constrained simulations of the Local Supercluster
predict high column densities of \ovi\ and \ovii\ in some directions
\citep{kravtsov02}.  It is thus likely that Galactic and extragalactic
phenomena both contribute to the local X-ray absorption, but the question
of which sightlines are dominated by which phenomena is, by and large,
unanswered.

Further complicating the issue is the presence of other gaseous components
of unknown origin.  \ion{H}{1} high--velocity clouds (HVCs) have a 
velocity distribution inconsistent with the Galactic rotation and therefore
are thought to be either neutral gas high in the Galactic halo (perhaps
from tidally stripped dwarf galaxies) or cooling, infalling gas from
the surrounding IGM.  Along many lines of sight studied with FUSE, 
high--velocity \ovi\ absorption lines (\ovihv) at velocities coincident with 
the \ion{H}{1} HVCs are seen, while in some other directions \ovihv\ is
present even in the absence of any significant \ion{H}{1} 21-cm emission
at the same velocity \citep{sembachetal03}.  The question of whether these
isolated \ovi\ HVCs arise in an extended, hot Galactic corona or at Local Group
scales -- and their relation to the $z=0$ X-ray absorption lines --
is a subject of continuing debate.

There is some evidence for such a Galactic corona 
\citep[discussed in detail by][and references therein]{sembach03}; indeed, 
a significant fraction of HVCs appear to exhibit
low--ionization absorption lines (such as C\,{\sc ii}--{\sc iv} and 
Si\,{\sc ii}--{\sc iv}), unlikely to arise in a low--density,
warm--hot photoionized medium \citep{collins05}.
However, \citet{nicastro03} showed that the velocity 
distribution of these unassociated \ovihv\ clouds is minimized in the
Local Group rest frame, indicating that their origin is extragalactic.
Additionally, W05 found that the \ovihv\ toward Mrk 421 cannot be 
associated with the \ovii\ along that sightline (assuming a single 
temperature/density phase).
However, the links between the \ovihv, \ion{H}{1} HVCs, and 
local \ovii/\oviii\ absorption in the context of the Galactic corona
and local WHIM are to a large degree unknown.  Determining the origin
of these components, and the relations between them, 
is a crucial step in our understanding of the ongoing process of galaxy 
formation

The nearby ($z=0.03$), X-ray bright Seyfert galaxy Mrk~279 lies in the
direction of the \ion{H}{1} HVC Complex C, thus providing a particularly
valuable background source
that can be used to study these gaseous components.  Here we report
on our analysis of deep \chandra\ and FUSE spectra of this object,
the detected UV and X-ray absorption, and the implications for gas in the
Galaxy and Local Group.

\section{Data Reduction and Measurements}
\subsection{\chandra}
Seven observations of Mrk 279 taken with the \chandra\ High Resolution
Camera Spectroscopic array (HRC-S) and Low Energy Transmission Grating (LETG),
all taken in May 2003 and totaling 340 ks of exposure time, were available
in the \chandra\ archive.  Each of these data sets was processed, and
instrument response files built, using the
standard data reduction threads for HRC-S/LETG\footnote{Available at  
\url{http://cxc.harvard.edu/threads/gspec.html}} with version 3.3 of 
the \chandra\ {\it Interactive Analysis of Observations (CIAO)} software and 
\chandra\ Calibration Database version 3.2.0.  Since the HRC-S does not
have sufficient energy resolution to distinguish LETG spectral orders, 
higher orders can increase the apparent flux at long 
wavelengths\footnote{See \url{http://cxc.harvard.edu/ciao/threads/hrcsletg\_orders/}}.
This effect was mitigated by including response files for orders $-6$ through 
$+6$ in our analysis; the inclusion of orders beyond these had 
an insignificant effect on the computed instrumental response.

The seven observations were coadded for a final (unbinned, with
$\Delta\lambda=0.0125$\,\AA) 
signal--to--noise ratio of ${\rm S/N}\sim 6.5$ near 22\AA.
We used the spectral fitting program {\it Sherpa} to fit a power law and
foreground Galactic absorption to the spectrum over $10-100$~\AA\ band 
(excluding the $49-57.5$~\AA\ and $60.5-67.5$~\AA\ chip gap regions).  The
relative Galactic abundances of carbon, nitrogen, oxygen, and neon were left
as free parameters in order to produce a better fit around the absorption
edges.  A power--law slope of $\Gamma=2.3$ and equivalent hydrogen column 
density of 
$(1.78\pm 0.03)\times 10^{20}$~cm$^{-2}$ is derived, agreeing reasonably 
well with the \citet{elvis89} value of $1.64\times 10^{20}$~cm$^{-2}$ near
this sightline.  The best--fit abundances of N and Ne were approximately
zero, oxygen equal to the solar value, and carbon 0.15 solar, though we 
re--emphasize
that these do not reflect actual Galactic abundances but rather provide 
a better fit near absorption edges where the calibration is uncertain.
A few weak, broad residuals remained afterward, probably
due to calibration uncertainties or source variability; these were corrected 
by including four broad Gaussians in the source model 
\citep[analogous to the technique described in][]{nicastro05a}. 

\begin{figure}  
\plottwo{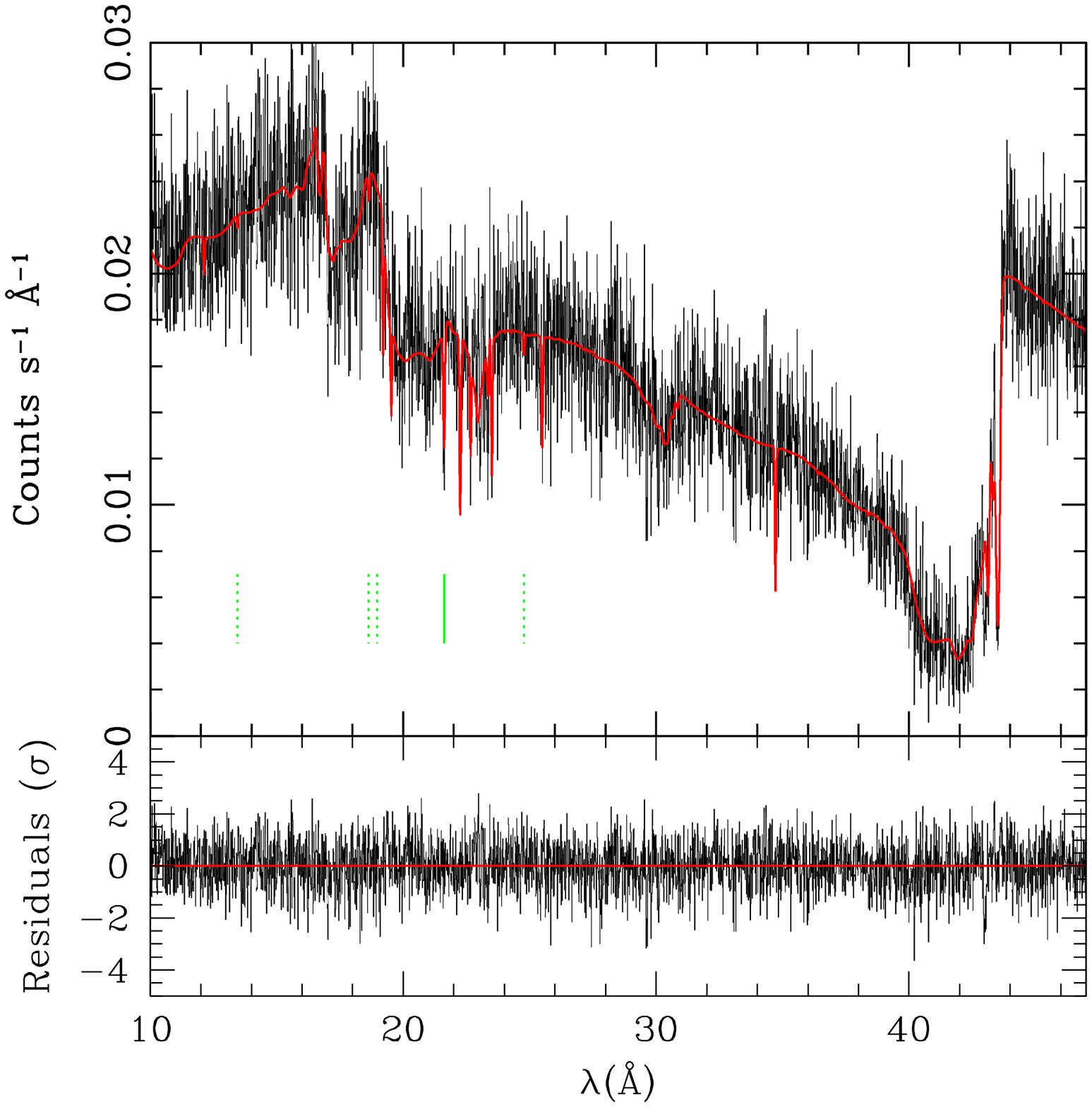}{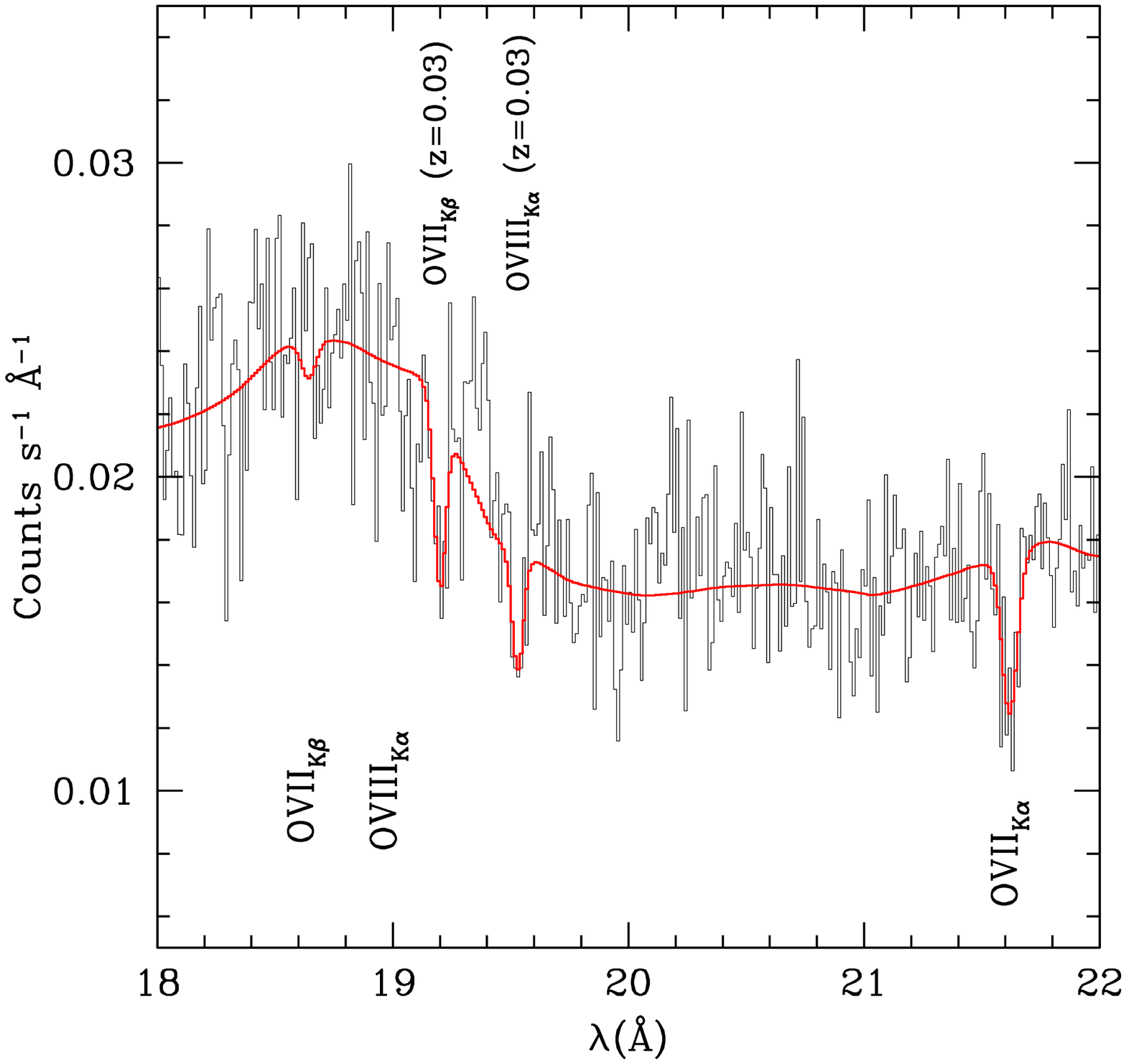}
\caption{\emph{Left:} 10--47\,\AA\ region of the coadded \chandra\ HRC--S/LETG 
spectrum of
Mrk~279 (top panel, black line) with the best--fit continuum model 
shown as the
red line, and residuals (bottom panel). The \ovii\ wavelength is marked
by the solid green line, and dotted green lines show the positions
of the measured upper limits listed in Table~1.  \emph{Right:} 18-22\,\AA\
region of the left panel showing in detail the $z=0$ \ovii\ K$\alpha$/K$\beta$ 
and \oviii\ K$\alpha$ regions.  Here, the \ovii\ K$\beta$ amplitude is 
constrained to the minimum (unsaturated) value, 
$W_\lambda({\rm K}\beta)=0.156\times W_\lambda({\rm K}\alpha)$.
\label{fig_chandraspec}}
\end{figure}

Once the continuum was established, we visually searched the spectrum 
in $\sim 3$~\AA\ windows for narrow (unresolved) 
absorption lines, fitting each one with a Gaussian.
Although several strong lines such as \cvi, \ovii, and \nvii\ 
are apparent at the AGN redshift ($z=0.03$), at $z=0$ ($v\la 700$\kms) 
only \ovii~K$\alpha$ $\lambda 21.602$ is 
unambiguously detected at $21.619\pm 0.009$\AA\ ($v=236\pm 125$\kms) 
with an equivalent width of 
$26.6\pm 6.2$~m\AA.  Upper limits were measured for the \ovii~K$\beta$ line
as well as several other ionic species of interest; these measurements
are listed in Table~\ref{tab_lines}.  Although they are included in the model
to improve consistency, the absorption intrinsic to Mrk~279 and the Galactic
interstellar \ion{O}{1} lines are not the focus of this work and will 
not be discussed further.  
The resulting fit and residuals are shown in Figure~\ref{fig_chandraspec};
with a reduced $\chi^2$ value of 0.89, the model appears to fit the data
quite well.

The \chandra\ HRC-S/LETG wavelength scale is uncertain for several reasons,
primarily because of non--linearities resulting from bad amplifiers on the HRC-S
detector\footnote{See \url{http://cxc.harvard.edu/cal/Letg/Corrlam/}}.
While the newly--released CIAO 3.3 software includes a routine to correct
these non--linearities and has reduced the dispersion
in wavelength errors to $\sim 6$\,m\AA\ at short wavelengths in calibration
spectra,
this routine is in the early stages of development and wavelengths of individual
emission and absorption lines may still be systematically skewed.
However, any systematic wavelength errors should not vary between observations
as long as the telescope pointing is nominal.  Furthermore, while serious
wavelength errors are known to occur around 18\AA, no bad amplifiers are
expected to significantly affect the dispersion relation near 21.6\AA\ 
(J.~J.~Drake and N.~Brickhouse, private communication).  

To check the absolute wavelength
scale near the \ovii\ line, we retrieved the nearest HRC--S/LETG calibration 
observation of the X-ray bright star Capella (observation 3675, taken on 2003
September 28) from the \chandra\ archive and 
reprocessed the data in exactly the same manner as the Mrk 279 data.
The wavelength of the strong \ovii\ emission line was found to be 
$21.606\pm 0.002$\AA\ or $56\pm 28$\kms, which is consistent with the 
$+30$\kms\ radial velocity of Capella as listed in the
SIMBAD database\footnote{\url{http://simbad.u-strasbg.fr/sim-fid.pl}}.
As a separate check, we reduced the Mrk~279 \chandra\ data both with
and without the wavelength correction routine; the difference in the measured
\ovii\ wavelength between the two was only 4\,m\AA, much lower than the
statistical error on the line position.
Thus, although the possibility of systematic wavelength errors must be
kept in mind, it appears as though such effects 
are insignificant compared to the statistical error on the measured
\ovii\ wavelength.

\subsection{\fuse \label{sec_meas_fuse}}
Mrk~279 was observed four times with FUSE between December 1999 and May 2003
with a total exposure time of 224~ks; all calibrated data from these 
observations were obtained through the Multimission Archive at STScI 
website.\footnote{\url{http://archive.stsci.edu/}}  To account for small
shifts in the \fuse\ wavelength scale during the observations, each of the
constituent exposures was cross--correlated over the $1030-1040$\AA\ range
and the relative positions of strong absorption lines were checked by
eye.  The data from 18 May 2002 were not of sufficient 
quality to reliably perform this wavelength calibration and were thus 
excluded.  The coadded, 
wavelength-shifted spectra from each observation were then cross--correlated
with each other, scaled so that their continuum intensities matched, combined
and rebinned by five pixels ($\sim 10$\kms) to produce a final spectrum
with ${\rm S/N}\sim 27$ near 1032\AA\ and an effective
exposure time of 177~ks.   

To account for possible systematic offsets, the absolute wavelength scale of 
the final spectrum was corrected following the method employed by W05: 
the Galactic \ion{Si}{2} $\lambda 1020.699$ and \ion{Ar}{1} $\lambda 1048.220$
absorption profiles were fit with multiple Gaussian components and the column 
density--weighted average velocities calculated.  These were found to be
$-70.0$\kms\ and $-65.5$\kms\ respectively, while the 
average velocity of the \citet{wakker03} multi--component fit to the Galactic
\ion{H}{1} toward the Mrk~279 sightline is $v\sim -37$\kms.  
Since these low--ionization lines
are expected to co-exist with the \ion{H}{1}, a $+30$\kms\ shift was applied
to the wavelength scale of the FUSE spectrum.  

\begin{figure}  
\plotone{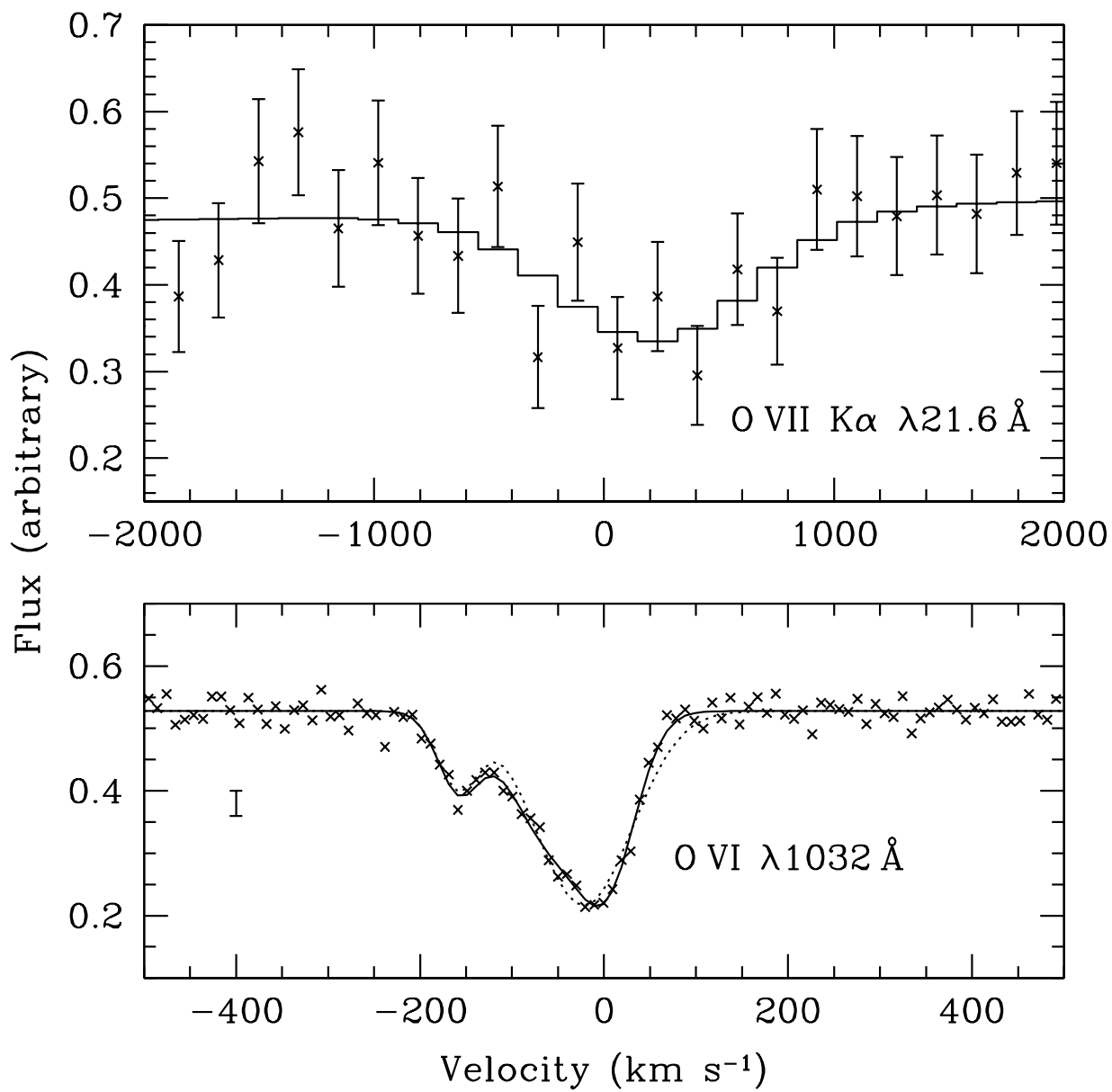}
\caption{Velocity relative to the local standard of rest of the 
local \ovii\ $\lambda 21.6$ (top panel) and \ovi\ $\lambda 1032$ absorption
lines, with the best-fit model plotted in each as the solid line; a 
representative error bar for the FUSE data points is shown at left.
Note the difference in scale between the two plots.
A single--Gaussian fit to the low--velocity \ovi\ absorption is also
shown as the dotted line.
The \ovii\ velocity is inconsistent with that of the \ovihv\ at the
$\sim 2.8\sigma$ level. \label{fig_velplot}}
\end{figure}

The final combined and calibrated FUSE spectrum shows strong \ovi\ 
absorption from the
Galactic thick disk at $v\sim 0$ and a weaker \ovi\ high velocity component
(\ovihv) at $v\sim -150$\kms (Figure~\ref{fig_velplot}).  Each of these
features was initially modeled with a single Gaussian; however, this provided
a poor fit for the thick--disk absorption due to an asymmetric blue wing on
the line profile; thus, the Galactic \ovi\ was instead fit with a broad
and a narrow Gaussian component (hereafter denoted \oviB\ and \oviN\ 
respectively).  Measured line parameters are also listed in 
Table~\ref{tab_lines}; note that the velocity of the \ovihv\ is tightly
constrained by this fit as $v=-159.9\pm 2.6$\kms, which is inconsistent
at the $2.8\sigma$ level with the \ovii\ velocity of $236\pm 125$\kms\ 
(assuming the HRC-S/LETG wavelength scale is correct),
indicating that these two components almost certainly are not related.

The total equivalent widths of the \ovi\ Galactic and 
high--velocity components are $262.5\pm 6.7$~m\AA\ and $43.6\pm 3.5$~m\AA\ 
respectively, in agreement within the estimated systematic errors of the
\citet{wakker03} measurements of $247\pm 8 \pm 25$~m\AA\ and 
$53\pm 6 \pm 17$~m\AA\ (where the first and second error values are 
statistical and systematic, respectively).  As was the case for the 
Mrk~421 sightline (W05), our fitted Galactic$+$HVC \ovi\ equivalent 
width agrees quite well with the \citet{wakker03} direct--integration value but
our measurement of the \ovihv\ strength is somewhat lower and Galactic 
absorption higher; this is most likely because our method better accounts
for blending between the Galactic and high--velocity \ovi\ in cases where
distinct absorption components are evident (as discussed in more detail
in W05).

\section{Analysis}
\subsection{Doppler Parameters and Column Densities \label{sec_cog}}
\begin{figure}   
\plotone{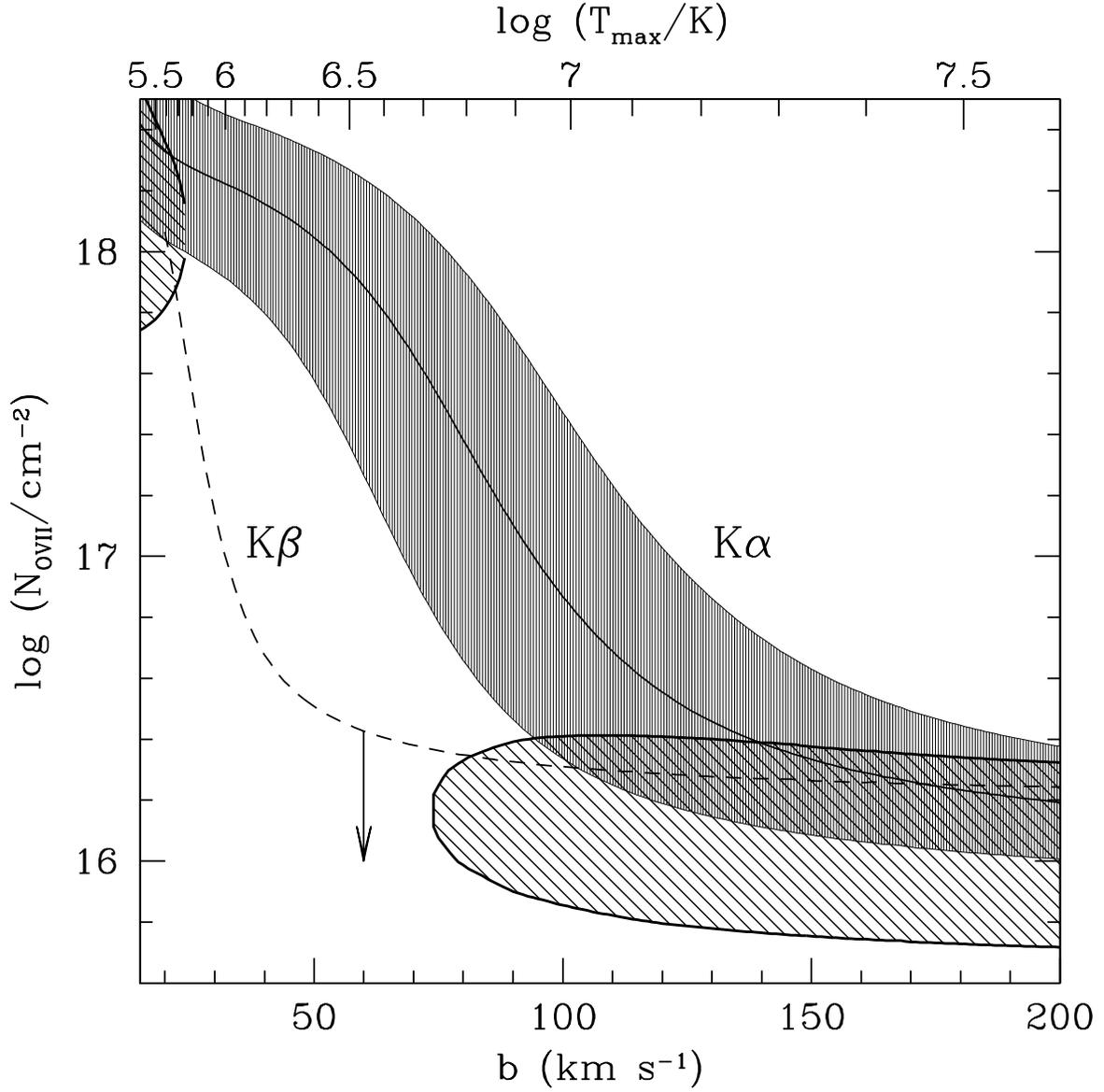}
\caption{Allowed values of $N_{\rm OVII}$ and $b$ given the measured 
\ovii~K$\alpha$ equivalent width and $1\sigma$ errors (shaded region) and
\ovii~K$\beta$ $2\sigma$ upper limit (dashed line).  Values of $\novii$ and 
$b$ for
which the two measurements are consistent (within 95\% confidence) 
are denoted by the hatched region. \label{fig_nhb_ovii}}
\end{figure}

In order to derive physical properties of the observed absorption, it is 
necessary to first convert the measured equivalent widths into ionic column 
densities $N_i$.  This requires knowledge of the Doppler parameter $b$,
since at a fixed column density $W_\lambda$ decreases for lower values of $b$.
However, the nominal LETG resolution of $0.05$\AA\ ($\sim 700$\kms\ at
21\AA) precludes direct measurement of the \ovii\ line width.  A method
similar to that employed in W05 is thus used to place limits on the 
\ovii\ Doppler parameter using the measured equivalent width of the
\ovii~K$\alpha$ line and the upper limit on \ovii~K$\beta$.  
Equivalent widths and FWHM values for each transition were calculated 
(assuming a Voigt absorption profile) over a grid of
$\novii$ and $b$, and the tracks in the $\novii-b$ plane 
consistent with each measured 
$2\sigma$ equivalent width limit are plotted in Figure~\ref{fig_nhb_ovii}.

Determining the ranges of $\novii$ and $b$ for which the measured
column densities are 
consistent with each other (i.e. the area over which the K$\alpha$ and
K$\beta$ tracks ``overlap'') is more complicated.  In the Mrk~421
data the first three lines in the \ovii\ K--series were individually detected
at $\ge 3\sigma$ confidence.  In that case, the differences between
pairs of predicted $\novii$ values (and the joint errors on the differences)
were calculable, allowing easy determination of the region over which the
three transitions predicted consistent $\novii$ and $b$ values.
In the case of Mrk~279, however, this same method cannot be used because
while the \ovii\ K$\alpha$ line has been detected, only an upper limit
is available for the K$\beta$ line.

However, since the absorption line properties for various column densities
and Doppler parameters are known, limits on these quantities 
can be determined using the \chandra\ spectrum itself.  For each point
in the $\novii-b$ plane, \ovii\ K$\alpha$ and K$\beta$ absorption
lines with the calculated $W_\lambda$ and FWHM values were added to 
the best--fit continuum model, and the $\chi^2$ statistic calculated using
the ``goodness'' command in Sherpa.  Over the calculated parameter ranges,
the minimum $\chi^2$ point was on the $b=200$\,\kms\ boundary.  
Since the K$\beta$ line was completely undetected (with a best--fit amplitude 
near zero), the $\chi^2$ value asymptotically approaches a minimum
as $b$ increases.  We thus assumed a minimum $\chi^2$ value from a fit
with a fixed unsaturated line ratio 
($W_\lambda(K\beta)=0.15*W_\lambda(K\alpha)$) and calculated
$\Delta\chi^2=\chi^2(\novii,b)-\chi_{\rm min}^2$ for every point.  
The 95\% confidence interval ($\Delta\chi^2 < 6$) determined with this 
method is shown in
Figure~\ref{fig_nhb_ovii}; at this confidence
level all Doppler parameters between $24<b<74$\kms\ are ruled out.

\begin{figure}  
\plotone{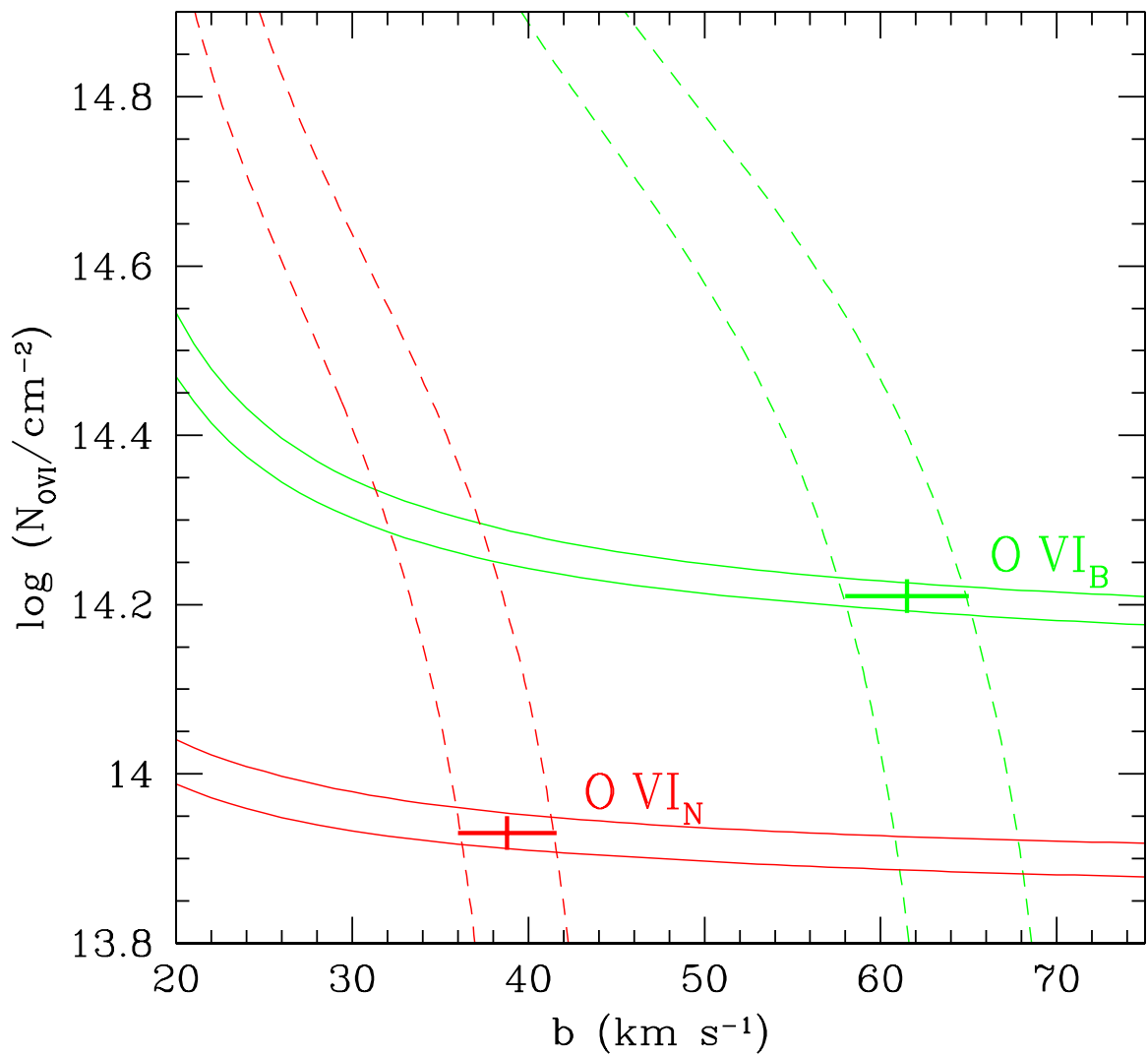}
\caption{Contours of constant equivalent width (solid) and FWHM (dashed)
for the low--velocity \ovi\ absorption, at the $1\sigma$ level.
Red contours are derived from the narrow low--velocity component and green 
contours from the broad component.  The inferred $1\sigma$ values of $\novi$
and $b$, as listed in Table~\ref{tab_lines}, are shown as crosses.  
\label{fig_nhb_ovi}}
\end{figure}

Although these curve--of--growth diagnostics can in principle also be applied 
to the UV \ovi\ 
$\lambda\lambda 1032,1038$ absorption doublet, it is typically more
difficult because few Galactic \ovi\ sightlines are sufficiently saturated
to significantly affect the doublet line ratio; additionally,
the close proximity of the 1037\AA\ line to other Galactic
absorption features makes deblending difficult \citep[W05;][]{wakker03}.
However, since the \ovi\ $\lambda 1032$ line is fully resolved in the FUSE 
spectrum, the measured line width is highly sensitive to $b$ while the
equivalent width traces $N_{\rm OVI}$.  While $W_\lambda$ decreases 
for a saturated line of a fixed column density, the observed FWHM increases
from the unsaturated value of ${\rm FWHM}=1.665b$.  To account for these 
saturation effects, we computed \ovi\ $\lambda 1032$ equivalent widths and
FWHM values over a grid of $N_{\rm OVI}$ and $b$.  Regions for which the
FWHM and $W_\lambda$ measurements are consistent with predicted values
are shown for both \oviB\ and \oviN\ in Figure~\ref{fig_nhb_ovi}; 
since the FWHM and $W_\lambda$ 
regions overlap nearly orthogonally, strong constraints are placed on
the column density and velocity dispersion of the \ovi--bearing gas.
None of the \ovi\ broad, narrow, or high--velocity components are 
significantly saturated, with derived Doppler parameters of $61.5\pm 3.5$\kms,
$38.8\pm 2.8$\kms, and $32.0\pm 4.6$\kms\ ($1\sigma$ errors) respectively.

All \ovi\ component Doppler parameters are inconsistent with the upper
allowed range ($b>74$\kms) determined from the \ovii\ ratios, and 
only the \ovihv\ is marginally consistent with the lower \ovii\ 
range ($b<24$\kms).  This indicates that the majority of the $v\sim 0$
\ovi\ cannot originate in the same phase as the observed \ovii\ absorption. 
Although at first glance it appears possible for the \ovihv\
and \ovii\ to coexist, the velocity separation of the \ovii\ and \ovihv\ 
(as discussed in \S\ref{sec_meas_fuse}) makes this interpretation
unlikely; additionally, the large \ovii/\ovihv\ column density ratio 
($\novii\ga 17.7$; Figure~\ref{fig_nhb_ovii}) requires high temperatures 
which are in turn ruled out by the low $b$ value (discussed further in the
following section).  It is also possible that the thick--disk \ovi\ consists of
only one component with a non--Gaussian shape, in which case the actual 
velocity dispersion could be larger.  Approximating the low--velocity \ovi\
with a single Gaussian component yields $b\sim 75$\kms --- barely consistent
with the 95\% lower limit on the \ovii\ Doppler parameter.  
We thus conclude that the \ovii\ most likely does not
coexist with any of the measured \ovi\ components.

\subsection{Temperature and Density Diagnostics}
The derived ionic column density ratios (and upper limits thereupon) can 
be used to determine the physical state of the absorbing medium.  Since the
extent, and hence the density, of the absorber is unknown, we cannot assume
a priori whether or not photoionization from the Galactic or extragalactic
background plays a significant role in the ionization balance of the 
medium.  Specifically, in a higher--density, hot medium (such as a Galactic
corona), collisional ionization would be the dominant process by far,
while in a low--density WHIM scenario photoionization is expected to play
a significant role.  Including the effects of photoionization along 
with collisional ionization is crucial in order to most accurately determine 
the state of the gas \citep[cf.][W05]{nicastro02,mathur03}.

We used version 05.07 of the ionization balance code Cloudy 
\citep{ferland98}  to calculate ionic abundances for all measured elements
over a grid of $T=10^{4.5}-10^{7.4}$\,K and ionization parameters 
$\log(U)=-6.3$~to~$0.7$ (where $U=n_\gamma/n_e$ is the ratio of the
number densities of ionizing photons and electrons in the plasma), with
a step size of 0.1 dex in both $\log T$ and $log U$.
The \citet{sternberg02} fit to the metagalactic radiation field was
assumed; this is based on the observed background from infrared to X-rays
(except for the unobservable radiation near the Lyman limit, which
is taken from the theoretical model of \citet{haardt96}).  The normalization
of this background corresponds to
electron densities of $n_e=10^{-7}-1$\,cm$^{-3}$ over the calculated $\log(U)$
range --- i.e., $\log(n_e)=-6.3-\log(U)$.  

As equivalent widths of different transitions from the same ion can be
used to place limits on the absorber's velocity dispersion and column density
(\S\ref{sec_cog}), any measured column
density \emph{ratio} defines a track in the $\log T-\log U$ plane. The overlap 
between two or more such
tracks (derived from different ion ratios) can then be used to place
constraints on the gas temperature and density.  Although in principle
the ratio between any two ionic species can be employed, column
density ratios between different ions of the same atomic species 
are metallicity--independent and hence produce the strongest diagnostics.
Since only \ovii\ is strongly detected at $z=0$ in the Mrk~279 \chandra\
spectrum and \ovi, \ovii, and \oviii\ together provide strong temperature
and density constraints \citep{mathur03}, we focus primarily on these ions.

A limit on $\noviii$ can be easily determined from the \oviii\ equivalent width
upper limit and assuming the \ovii\ Doppler parameter; however, including
the \ovi\ column density is less straightforward due to the presence of
multiple components.  Since the \ovii\ Doppler parameter is inconsistent
with both of the broad $v\sim 0$ \ovi\ components, there are two likely
possibilities: (1) the \ovii\ Doppler parameter is actually in the lower
range ($b< 21$\kms) and the \ovii\ is associated with the \ovihv, 
or (2) the \ovi\ absorption produced by the \ovii--bearing gas is too
weak to be detected in our FUSE spectrum, so only an upper limit on $\novi$
can be used in this analysis.  The former case is highly unlikely ---
not only is the centroid of the \ovii\ line inconsistent with the velocity
of the \ovihv, but such a low Doppler parameter requires extremely
high \ovii\ column densities ($\novii\sim 10^{18}$\,cm$^{-2}$).  This in
turn produces an extremely large $\novii/\novi$ ratio which requires
high temperatures ($T>10^{7.4}$\,K, the upper limit of our calculation).
Since the \ovihv\ Doppler parameter implies a maximum temperature
of $T_{\rm max}\la 10^6$\,K, such an association appears impossible.

\begin{figure} 
\plottwo{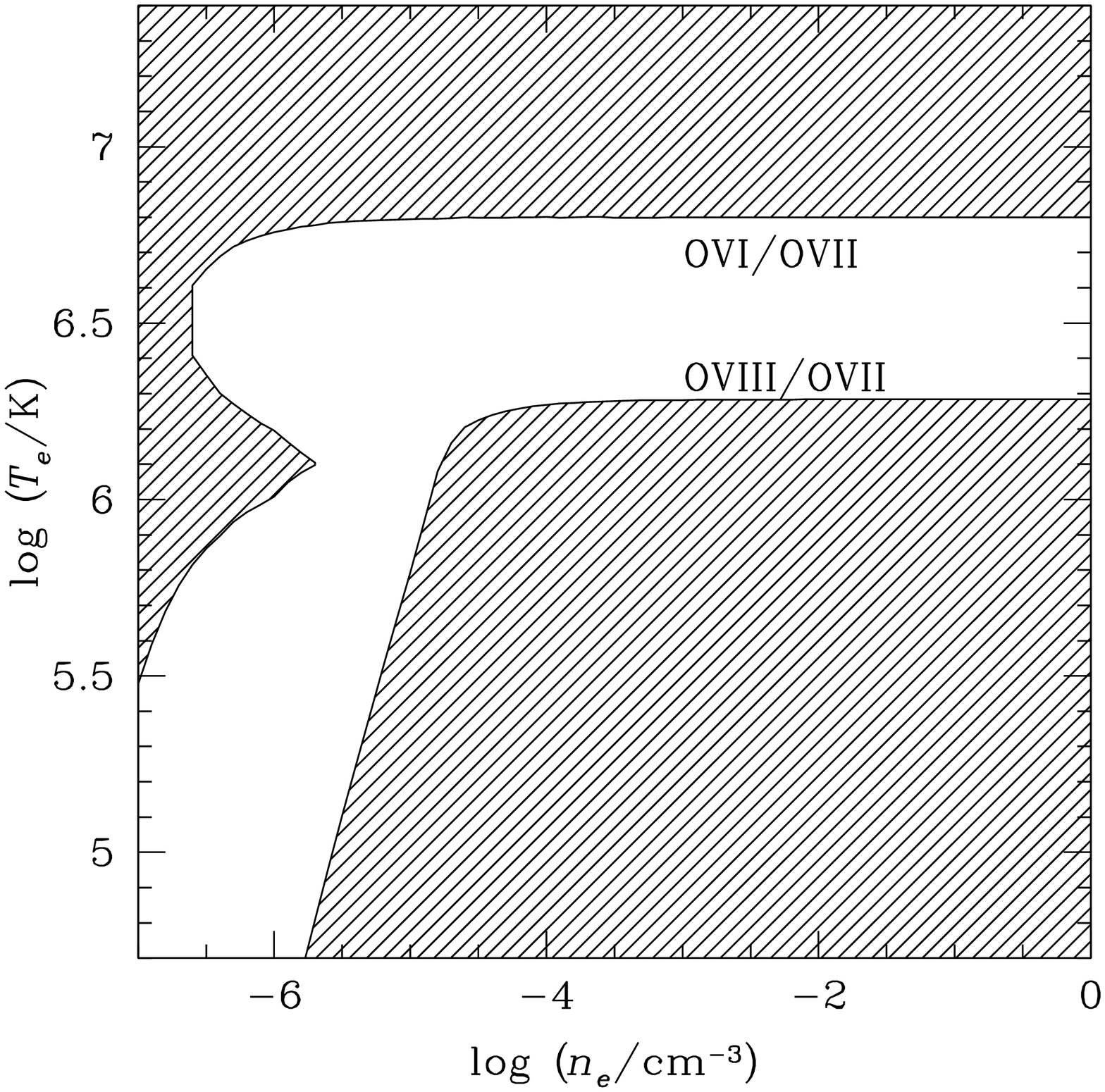}{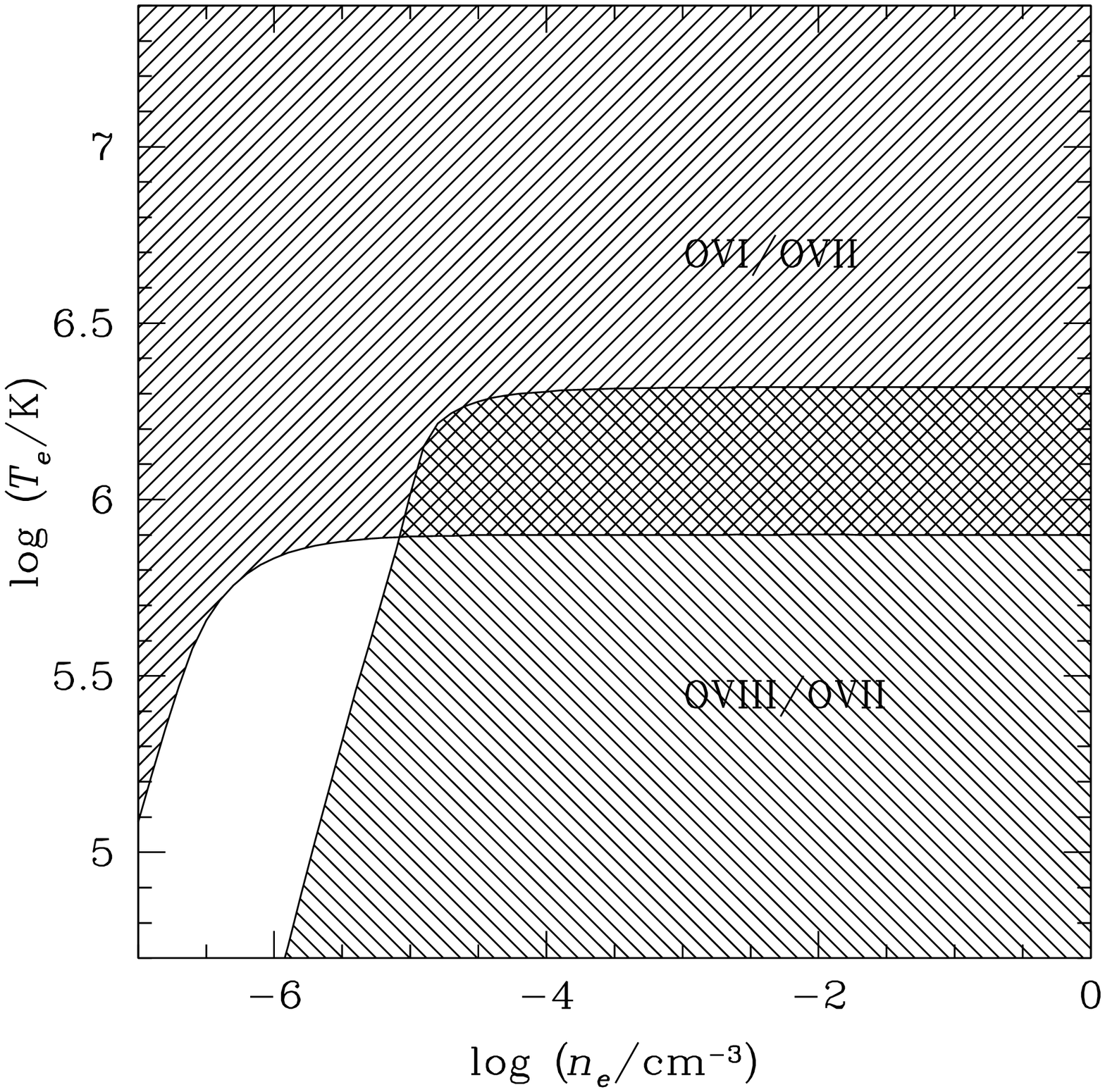}
\caption{Regions of consistency in the temperature--density plane for 
the $2\sigma$ \oviii/\ovii\ and
\ovi/\ovii\ column density ratio upper limits.  Here the \ovi\ upper limit
is calculated from a putative \ovi\ absorption line with
$b=100$\kms\ (left) and $b=200$\kms\ (right)  superposed on the
Galactic \ovi\ absorption.  While a consistent solution cannot be found for low
velocity dispersions, at higher values of $b$ the contours begin to overlap.
\label{fig_temp_dens}}
\end{figure}

If, instead, the associated \ovi\ absorption is too weak to be detected,
then this absorption most likely takes the form of a broad ($b\ga 80$\kms,
from the \ovii\ $b$ limit) absorption line superposed on the $v\sim 0$ 
\ovi.  Such a line was included in the FUSE spectrum fit and $2\sigma$
upper limits calculated for a velocity dispersions of 100 and 200\kms.
The temperatures and densities consistent with the \oviii/\ovii\ and
\ovi/\ovii\ upper limits for both values of $b$ are shown in 
Figure~\ref{fig_temp_dens}.  The \oviii/\ovii\ ratio sets an upper limit
of $\log T\la 6.3$ and a minimum density of $10^{-6}$\,cm$^{-3}$ in both
cases.  For $b=100$\kms\ the limits set by $\novi/\novii$ are inconsistent
with the $\noviii/\novii$ ratio, but for $b=200$\kms\ the contours begin to 
overlap with
$5.9<\log T<6.3$ and $\log n_e>-5.1$.  Thus, if the \ovii\ is associated
with a broad undetected \ovi\ component, a large Doppler parameter is
required to reconcile the oxygen ion ratios.  This, along with the inferred
temperature and density limits, are both consistent with expectations
for the local WHIM, though the large velocity dispersion compared to the
upper temperature limit derived from the \oviii/\ovii\ ratio indicates
that the line broadening is primarily nonthermal.

Although a similar analysis can be performed with the 
\nvii/\ovii\ and \neix/\ovii\ ratios, the derived limits are in all
cases weaker than those set by the \oviii/\ovii\ upper limit and have
been excluded from Figure~\ref{fig_temp_dens} for the sake of clarity.

\subsection{The AGN Warm Absorber}
The \chandra\ spectrum of Mrk~279 shows strong \ovii, \oviii, \cvi, 
and \nvii\ absorption at a redshift consistent with the AGN ($z=0.03$).
It is thus possible that the \ovii\ absorption line at $z=0$ is
contaminated, either by \ovii\ outflowing from Mrk~279 or from
another absorption line at $z=0.03$.  The former scenario is probably not the
case since this would require both an unlikely coincidence of the
outflow velocity with the AGN redshift
($v \sim 9000$\kms).  To check the latter case, we used the PHASE
model \citep[described by][]{krongold03} to fit the intrinsic absorption in a
self--consistent manner.   With this fit we found that the redshifted \nvii\
K$\beta$ line falls at 21.5\AA, near the \ovii\ K$\alpha$ rest wavelength
but well outside the line profile, and is weak enough that its effect
on the \ovii\ $K\alpha$ equivalent width measurement is most likely
negligible.  No other warm--absorber lines are expected near 21.6\AA\
for outflow velocities between zero and 9000\,\kms; thus, the $z=0$
\ovii\ measurement is unlikely to be contaminated by any lines from the
warm absorber.  The details of the warm absorber model are the subject
of a forthcoming paper (D.~L.~Fields et al., in preparation) and will
not be discussed further here.

\section{Discussion}
\subsection{Comparison to the Mrk 421 Sightline}
Although the quality of the Mrk~279 \chandra\ spectrum is far lower than that
of Mrk~421 (W05), the differences between the local absorption seen along 
the two 
lines of sight are striking.  While the velocity of the Mrk~421 \ovii\ 
absorption is near zero and thus cannot be distinguished from the 
low-- and high--velocity \ovi\ seen in the spectrum, the Mrk~279 \ovii\
and \ovihv\ velocities are significantly different.  Furthermore, the
derived Doppler parameters of the \ovii\ absorption along these 
sightlines --- $24<b<55$\kms\ and $b>74$\kms\ for Mrk~421 and Mrk~279
respectively --- differ substantially.  In both cases association of the
\ovii\ with any \ovi\ component is ruled out and the derived temperature,
density, and column density limits are consistent with each other (though
with large errors).  However, the strong discrepancy between the velocity
dispersions of the two absorbers suggests that their origins may differ.
Such a difference in origin may not be surprising, given the large
($\Delta l\sim 60^\circ$) separation between the two sightlines.

\subsection{Origin of the Absorption}
The unique properties of the absorption components along the Mrk~279 
sightline provide some tantalizing clues as to the origin of the local host
gas.  Taking all components into 
account, this sightline exhibits high negative--velocity \ion{H}{1}
emission (Complex C), \ovi\ absorption at a similar velocity, and broad, 
possibly redshifted \ovii\ absorption.  If Complex C is indeed
nearby WHIM gas that has cooled and is falling onto the Galaxy 
\citep[e.g.][]{miville05}, then its presence could indicate the presence of a
large--scale WHIM filament in the same direction.  In this interpretation,
the large nonthermal Doppler parameter of the \ovii\ absorption
could be a result of velocity shear, due either to the Hubble expansion 
over a scale of $\sim 3$\,Mpc (with a corresponding density of $10^{-5}-
10^{-4.5}$\,cm$^{-3}$, assuming $b=200$\kms, pure Hubble broadening and a 
metallicity of $0.1-0.3\times$ solar) or the natural velocity distribution
expected from infalling gas, or a combination of both.  The
negative--velocity \ion{H}{1} would then be gas that has ``broken off''
from the filament and is now falling onto the Galaxy, with the \ovi\ at the
same velocity representing the cooling component of the gas.  The velocity
of the local standard of rest is approximately perpendicular to the 
CMB rest frame in this direction, so no significant additional velocity shifts
are expected in an IGM scenario.  

Such a picture is consistent with the general picture of galaxy formation
and accretion of gas onto galaxies from the IGM, and the 
temperatures and densities inferred from the X-ray absorption are consistent
with those expected from the WHIM.  However, unlike the Mrk~421
sightline, the simulations of \citet{kravtsov02} do not predict high
column densities of \ovii\ in this direction (though this may be due to 
the limited resolution of the simulations).  Furthermore, aside from 
Complex C there are no known structures in this direction that might
indicate the presence of a local filament --- \citet{wakker03} note
that the Canes Venatici Galaxy Grouping is centered on this sightline
at $v\sim 2400$\kms, but this is far higher than the velocity
of the \ovii\ absorption.

Given these caveats and the large uncertainties on the X-ray measurements
(indeed, only
one $z=0$ absorption line has been strongly detected), the absorption
could also originate locally in hot Galactic halo or coronal gas.  
As discussed previously, such an origin would require the \ovii\ to
be a completely separate component from any of the other observed
components (\ion{H}{1}, low-- and high--velocity \ovi) along this line
of sight.  Of course, there is also the possibility that the \ovii\ 
absorption actually consists of multiple unresolved components, in which
case a multiphase solution may reconcile the discrepancy with the \ovi\ 
absorption.  All line ratio calculations were performed under the
assumption of ionization equilibrium, so nonequilibrium scenarios could
provide substantively different predictions as well.
More detailed modeling and simulations of both the Galactic
and IGM gas distributions will be necessary to determine which scenario
is most likely, and most consistent with the data.

\section{Conclusions}
Long--duration \chandra\ grating observations of the bright AGN Mrk~279
reveal the presence of strong \ovii\ K$\alpha$ absorption at a redshift 
consistent
with zero.  A FUSE spectrum of the same source shows several additional
\ovi\ components at velocities near zero.  Through kinematic, 
curve--of--growth, and ionization balance modeling, we conclude the 
following:
\begin{enumerate}
\item{A direct $\chi^2$ analysis of the \chandra\ spectrum 
coupled with absorption line models constrains the Doppler parameter of the
\ovii\ absorption to be $b>74$\kms\ and $b<24$\kms.
This latter range is unlikely due to the extremely high \ovii\ column
densities required to produce the strong absorption feature.}
\item{The \ovii\ Doppler parameter limits are inconsistent with the
measured $b$ values for any of the $v\sim 0$ \ovi\ absorption
components.  Additionally, the centroid of the \ovii\ K$\alpha$ line
is inconsistent (at the $2.5\sigma$ level) with that of the \ovihv, indicating
that the \ovii\ is not associated with \emph{any} local \ovi\ component.}
\item{If the \ovii\ absorption is associated with a broad, undetected
\ovi\ absorption line, then a large Doppler parameter ($b\sim 200$\kms)
is required to provide a single--phase solution for the \ovi, \ovii, and
\oviii\ column densities.  This large value of $b$ could be a result of
either microturbulence, velocity shear from infalling gas, or broadening 
due to the Hubble expansion over a path length of a few Mpc.  If the line
is purely Hubble--broadened, at $b=200$\kms\ a pathlength of 3\,Mpc and
density of $\log n\sim -5$ is implied (assuming an oxygen
abundance of 0.3 times solar).}
\item{The large velocity dispersion, possible redshift, and lack of
association with any Galactic absorption components (as well as the
proximity of HVC Complex C) indicates that this X-ray absorption may
be from a large--scale nearby WHIM filament; however, a Galactic
corona origin cannot be ruled out with the current data.}
\end{enumerate}



\acknowledgments

The authors thank the \chandra\ and FUSE teams for their efforts on these
superb missions, and are in particular grateful to Martin Elvis for his
comments on a draft of this paper and Jeremy Drake and Nancy Brickhouse
for helpful discussions regarding the \chandra\ calibration.  We also
appreciate the anonymous referee's helpful comments on the manuscript.
Ionization balance calculations were performed with 
version 05.07 of Cloudy, last described by \citet{ferland98}.
This research has been supported by \chandra\ award AR5-6017X issued by the
\chandra\ X-ray Observatory Center, which is operated by the Smithsonian
Astrophysical Observatory for and on behalf of the NASA under contract
NAS8-39073.  RJW derives additional support from an Ohio State University
Presidential Fellowship.




\clearpage

\begin{deluxetable}{lccccccc}
\tabletypesize{\footnotesize}
\tablecolumns{8}
\tablewidth{500pt}
\tablecaption{Observed $z\sim 0$ absorption lines \label{tab_lines}}
\tablehead{
\colhead{ID} &
\colhead{$\lambda_{\rm rest}$\tablenotemark{a}} &
\colhead{$\lambda_{\rm obs}$\tablenotemark{b}} &
\colhead{$\Delta v_{\rm FWHM}$} &
\colhead{$v_{\rm obs}$} &
\colhead{$W_\lambda$\tablenotemark{c}} &
\colhead{$\log N_i$\tablenotemark{c,d}} &
\colhead{Note}
\\
\colhead{} &
\colhead{(\AA)} &
\colhead{(\AA)} &
\colhead{(\kms)} &
\colhead{(\kms)} &
\colhead{(m\AA)} &
\colhead{} &
\colhead{}
}

\startdata
X-ray: \\
\hline\hline
\ion{O}{7} K$\alpha$ &21.602 &$21.619\pm 0.009$ &$600^{+400}_{-600}$ &$236\pm 125$ &$26.6\pm 6.2$ &$16.19\pm 0.19$ &1\\
\ion{O}{7} K$\beta$  &18.629 &18.629 &\nodata &\nodata &$<7.0$ &$<16.24$ &\\
\ion{O}{8} K$\alpha$ &18.969 &18.969 &\nodata &\nodata &$<6.5$ &$<15.72$ &\\
\ion{N}{7} K$\alpha$ &24.781 &24.781 &\nodata &\nodata &$<7.0$ &$<15.51$ &\\
\ion{Ne}{9} K$\alpha$&13.447 &13.447 &\nodata &\nodata &$<7.3$ &$<15.88$ &\\ 
\hline
UV: \\
\hline\hline
\oviB &1031.926 &$1031.75\pm 0.01$ &$112.3\pm 5.8$ &$-50.8\pm 3.5$ &$169.9\pm 5.5$ &$14.21\pm 0.02$ &\\
\oviN &1031.926 &$1031.95\pm 0.01$ &$69.9\pm 4.4$  &$6.7\pm 1.8$   &$92.6\pm 3.9$ &$13.93\pm 0.02$  &\\
\ovihv &1031.926 &$1031.38\pm 0.01$ &$53.3\pm 7.6$ &$-159.9\pm 2.6$ &$43.6\pm 3.5$ &$13.58\pm 0.04$\\
\enddata
\tablenotetext{a}{Rest wavelengths taken from \citet{verner96}.}
\tablenotetext{b}{In the cases where upper limits were found, the line 
positions were frozen to the rest wavelengths.}
\tablenotetext{c}{Error bars are $1\sigma$; upper limits are $2\sigma$.}
\tablenotetext{d}{Column densities for X-ray lines are calculated assuming
$b=200$\kms; for UV lines the measured $b$ values are used.}
\tablecomments{
(1) The column density given here for the \ovii\ K$\alpha$ line is from the
equivalent width measurement assuming $b=200$\kms, not the $\chi^2$
method described in \S\ref{sec_cog}.
}
\end{deluxetable}

\end{document}